\documentclass[fleqn,10pt]{revtex4}

\usepackage{graphicx}
\usepackage{amstext}
\usepackage{color}
\usepackage{ulem}
\usepackage{amssymb, amsfonts}
\usepackage{textcomp, float, bbm}
\usepackage{soul}
\usepackage{hyperref}
\usepackage{lscape}
\usepackage[mathscr]{euscript}
\usepackage{xcolor} 

\newcommand{\ket}[1]{\left| #1 \right\rangle}                                   
\newcommand{\bra}[1]{\left\langle #1 \right|}                                   
\begin{document}
\title{Experimental simulation of decoherence in photonics qudits}

\author{B. Marques$^{1,2,*}$, A. A. Matoso$^{1}$, W. M. Pimenta$^{1}$, A. J. Guti\'errez-Esparza$^{1}$, M. F. Santos$^{1}$, and S. P\'adua$^{1}$}

\address{$^{1}$Departamento de F\'isica, Universidade Federal de Minas Gerais,caixa postal 702, 30123-970, Belo Horizonte, MG - Brazil}
\address{$^{2}$Department of Physics, Stockholm University, S-10691 Stockholm, Sweden}

\begin{abstract}
We experimentally perform the simulation of open quantum dynamics in single-qudit systems. Using a spatial light modulator as a dissipative optical device, we implement dissipative-dynamical maps onto qudits encoded in the transverse momentum of spontaneous parametric down-converted photon pairs. We show a well-controlled technique to prepare entangled qudits states as well as to implement dissipative local measurements; the latter realize two specific dynamics: dephasing and amplitude damping. Our work represents a new analogy-dynamical experiment for simulating an open quantum system.

\end{abstract}
\maketitle


\section*{\label{Introduction}Introduction}

In most cases, when a quantum system interacts with its environment it undergoes decoherence \cite{DECOERENCIA1}, to wit, the system-environment interaction ``spoils'' the state of the system by decreasing its capacity for quantum interference, which is essential for standard quantum information processing. 
Decoherence is so far one of the major obstacles for implementing quantum computation processes in real systems. Despite such nuisance, recent works have shown procedures to manipulate the system-environment interaction or the information leaked to the environment in suitable ways depending on the specific goal: \textit{e.g.}, estimation of quantum noise \cite{chiuri2011experimental}, protection of coherence and/or entanglement \cite{FRANCAANDRE}, universal quantum computation \cite{FRANCATERRA2012}, quantification of entanglement \cite{QUANTIFY1,QUANTIFY2,QUANTIFY3} and entanglement concentration \cite{BRENO1}.

An interesting study for exploring quantum devices is the experimental simulation of complex dynamics on controllable quantum systems of simple implementation \cite{RevModPhys.86.153}. 
These simulations allow for a better control and, therefore, understanding of the details leading to decoherence as well as the mechanisms underneath the system-environment exchange of excitation and/or information. Recent works on dynamics simulations have been performed in diverse quantum systems such as optical interferometers with polarization-entangled photon pairs generated by spontaneous parametric down-conversion (SPDC) \cite{OPENIMPLEMENTATION3,OPENIMPLEMENTATION4,OPENIMPLEMENTATION5,OPENIMPLEMENTATION000}, spin-$\frac{1}{2}$ nuclear states of carbon atoms accessed by magnetic nuclear resonance \cite{OPENIMPLEMENTATION2}, and trapped ions \cite{OPENIMPLEMENTATION1, OPENIMPLEMENTATIONBLATT, OPENIMPLEMENTATIONION20,OPENIMPLEMENTATIONION30}. 

In particular, the simulation of open system dynamics for qubits (two-level quantum systems) has already been observed in many different experiments and has been connected to the observation of phenomena such as entanglement sudden death \cite{OPENIMPLEMENTATION3} and Non-Markovian dynamics \cite{FABIO1, FABIO2, STEVE1}, to name a few. The extension of a similar analysis to qudits (d-level quantum systems) presents both a wider range of phenomena to observe and possible applications to explore ranging from full local protection of entanglement to dynamical precursors of entanglement sudden death that are not present in pair of qubits. However, the simulation of quantum open systems in qudits is not as easy to implement as in qubits which justifies the rarity of such results in the literature.

In this paper we report an experimental technique for simulating decoherence in the the dynamics of a qudit. Our qudits are encoded in the transverse component of the linear momentum of photon pairs generated by SPDC. The quantum system is defined in terms of the path entangled photons, namely, down-converted photons propagated through paths outlined by optical diffractive elements (multi-slits) \cite{TRANSVERSAL2}. In particular, we simulate two types of decoherence mechanisms namely dephasing \cite{DEFASAGEM} and amplitude damping \cite{AMPLITUDE}, by means of a spatial light modulator (SLM) employed to implement operations on the qudit states. A wide variety of applications extend the use of SLMs for controlled manipulation of photonic quantum systems encoded, for instance, in polarization \cite{OPENIMPLEMENTATION6},  in orbital angular momentum \cite{CONCORT1,BELLORBITAL} or in transverse momenta of the photons \cite{DEUTSCHBRENO,T2QUBITS,QUDITWITNESS4, T1QUBIT, SLMTRANSVERSAL2, SLMTRANSVERSAL1, SLMTRANSVERSAL3}. 

This article is organized as follows: in section~\ref{Theory} we summarize the concepts of open system dynamics; the state preparation and the dephasing implementation are discussed in sections \ref{setup} and \ref{Tdephasing}, respectively; section~\ref{Tamplitude} presents the amplitude damping implementation. We summarize and conclude the article in section~\ref{Summary}.

\section{\label{Theory}Open quantum systems}
In this section we present a brief review of the theory of open quantum systems in order to outline this work. A system ($S$) interacting with an environment ($E$) is described by the Hamiltonian \cite{OPENIMPLEMENTATION4}
\begin{equation}
H=H_S\otimes I_E + I_S \otimes H_E + H_{int},
\end{equation}
where $H_S$ and $H_E$ are the system and environment Hamiltonian operators, respectively; $H_{int}$ is the system-environment coupling Hamiltonian, and $I_S$ ($I_E$) is the system (environment) identity operator. We can describe only the system evolution by the equation of motion given by
\begin{equation}
\dot{\rho}_S=-\frac{i}{\hbar}\text{Tr}_E\left[H,\rho_{SE}\right],
\end{equation}
where $\rho_{SE}$ is the total ($S+E$) density operator. 

The system evolution for the system-environment coupling can always be expressed as an unitary evolution and the total density matrix can be written as
\begin{equation}
\rho_{SE}=U_{SE}(t)\rho_{SE}(0)U_{SE}^{\dagger}(t),
\end{equation}
where $U_{SE}(t)=exp(-iHt/\hbar)$ and $\rho_{SE}(0)$ is the density matrix of the initial state. In the particular case where we consider the initial state as a product state between the system and the environment, $\rho_{SE}(0)=\rho_S(0)\otimes \ket{0}{_E}\bra{0}{_E}$, the effective evolution of the system is given by:
\begin{eqnarray}
\rho_S(t)&=&Tr_E\left[U_{SE}(t)\rho_{SE}(0)U_{SE}^{\dagger}(t)\right],\\
&=&\sum_\epsilon ({_E}\bra{\epsilon} U_{SE}(t)\ket{0}_E )\rho_{SE} ({_E}\bra{0} U_{SE}^{\dagger}(t) \ket{\epsilon}_E) \nonumber,
\end{eqnarray}
where $\{\ket{\epsilon}\}$ form an orthonormal basis for the environment. This evolution can be expressed only in terms of operators acting on $S$ in the following form
\begin{equation}
\rho_S(t)=\sum_\epsilon K_\epsilon(t) \rho_S(0) K_\epsilon^{\dagger}(t),
\end{equation}
where the operators
\begin{equation}
K_\epsilon(t)= {_E}\bra{\epsilon} U_{SE}(t)\ket{0}_E
\end{equation}
are the so-called Kraus operators \cite{KRAUS1} and define a trace preserving positive map: $K_\epsilon^\dagger (t) K_\epsilon (t) >0$, $\sum_\epsilon K_\epsilon^\dagger (t) K_\epsilon (t) =1$, $Tr_S(\rho_S(t))=1$. Note that the Kraus operators are not uniquely defined because there are many bases for describing the environment. As a consequence, we deal with the equivalent operators from different sets, which originate different decompositions of the same resulting density matrix.

Under certain well established circumstances known as the Born-Markov approximations, the evolution of the system can also be expressed in terms of a time continuous Master equation, given by  \cite{MASTEREQUATION} 
\begin{equation}
\dot{\rho}_S=-\frac{i}{\hbar}\left[H_S,\rho_{S}\right] + \sum_{j=1}^{N^2-1}\gamma_j \left(A_j\rho_S A_j^\dagger-\frac{1}{2}\rho_S A_j^\dagger A_j - \frac{1}{2}A_j^\dagger A_j\rho_S\right),
\end{equation}
where $\rho_S$ is the system density operator, $A_j$ are the so-called Lindblad operators and $\gamma_j$ is a non-negative quantity which has dimensions of the inverse of time if $A_j$ is dimensionless. The first term on the right side of the master equation represents the unitary part of the dynamics generated by the Hamiltonian $H_S$. In this case, an intuitive set of Kraus operators is given by $K_0 = 1 - iH_Sdt - \sum_j \frac{\gamma_j dt}{2} A^\dagger_j A_j$ and $K_j = \sqrt{\gamma_j dt} A_j$ where $\gamma_j dt \ll 1$ and terms of the order of $dt^2$ are ignored. This so-called unravelling of the Master equation is associated to the Quantum Trajectories method where $K_0$ and $K_j$ are respectively known as the No-Jump and Jump operators. This method is connected both to an alternative way to calculated the evolution of the system on average as well as a direct way to infer its evolution at any single realization where a sequential measurement of the state of the environment is performed. 

In this work, as mentioned above, we focus on two types of open quantum system evolution: dephasing and amplitude-damping. In the following sub-sections we outline such evolutions.


\subsection{Dephasing}

In a dephasing dynamics the system loses coherence due to the system-environment interaction without any population exchange. This occurs when a noisy environment couples to a system \cite{DEFASAGEM}. We can describe this dynamics for a system with dimension $d$ using the Kraus operators
\begin{eqnarray}
K_d&=&I_d,\\
K_j&=&\sum_{i=0}^{d-1}e^{i\pi\delta_{ij}}\ket{i}\bra{i},
\end{eqnarray}
where $0\leq j\leq d-1$ and $I_d$ is the identity operator for a system of dimension $d$. The system evolution in a dephasing dynamics can be obtained from
\begin{equation}
\rho_S(\{p_i\})=\sum_{i=0}^{d}p_iK_i\rho_SK_i^\dagger,
\label{pspi}
\end{equation}
where $\{p_i\}$ is the set of time dependent parameters that represent the weight of each Kraus operator. Writing the system density operator in a matrix form such that $\bra{i}\rho\ket{j}=\rho_{ij}$, the dynamics for each matrix density element can be obtained. The diagonal elements are constant,
 \begin{equation}
 \rho'_{ii}=\rho_{ii}, \quad \forall~i,
 \end{equation}
 and the off-diagonals elements evolve as
 \begin{equation}
 \rho'_{ij}=(1-2p_i-2p_j)\rho_{ij} \quad \forall~i\neq j,
 \end{equation}
showing the system decoherence. The experimental implementation can be simplified if we consider the particular case $p_i=p/4$ for $0\leq j\leq D-1$, inducing a single-parameter dependence in the system evolution. Thus, the off diagonals elements are
\begin{equation}
\rho'_{ij}=(1-p)\rho_{ij} \quad \forall~i\neq j.
\end{equation}
Dephasing dynamics implementation is presented in section \ref{Tdephasing}.

\subsection{Amplitude Damping}
Damping dynamics represents the dissipative interaction between the system and its environment. A common example is the loss of photons from a cavity into a zero-temperature environment of electromagnetic-field modes~\cite{abdalla2014purity}. This dynamics can be described using the master equation
\begin{equation}
\dot{\rho}=2\gamma a\rho a^{\dagger} -\gamma \rho a^{\dagger}a-\gamma a^{\dagger}a\rho,
\end{equation}
where $a$ ($a^{\dagger}$) is the operator annihilation (creation) of a photon inside the cavity and $\gamma$ is the decay rate inside the cavity.

One approach to describe the system dynamics is based on the theory of quantum trajectories \cite{UNRAVELING0,UNRAVELING1,UNRAVELING2}, which consist in monitoring the system's environment. Environmental monitoring during a time interval $\{t,t+\delta t\}$ indicates whether or not a loss of excitation (a quantum jump) can occur. If no loss of excitation occurs, the system evolves without a quantum jump; thus, this evolution is given by

\begin{equation}
\rho_S(t+\delta t)=e^{-\frac{i}{\hbar}H_{eff}\delta t}\rho_S(t)e^{\frac{i}{\hbar}H_{eff}\delta t},
\end{equation}
where $H_{eff}=i\hbar\gamma a^{\dagger}a/2$. On the other hand, if the system loses an excitation, the system evolves with a quantum jump
\begin{equation}
\rho_s(t+\delta t)=\frac{\delta t}{\delta p}a\rho_s(t)a^{\dagger},
\end{equation}
where $\delta p=\delta t\gamma \text{Tr}[a\rho_s(t)a^{\dagger}]$ is the probability that a quantum jump occurs within the time interval. Note that, the higher the excitation in the cavity, the greater the chance of a quantum jump occurs, $\delta p=\delta t \gamma \text{Tr}[a\ket{N}\bra{N}a^{\dagger}]=\delta t \gamma N$, where $\ket{N}$ is the photon number state inside the cavity.

In Section \ref{Tamplitude} we demonstrate a partial dynamics of this open system: the no-jump trajectories. Such interesting dynamics occurs when the evolution exhibits no jump at all.

\section{Experimental setup}
\label{setup}
\begin{figure}[ht!]
\hspace{0.1cm}
\begin{center}
\includegraphics[width=9cm]{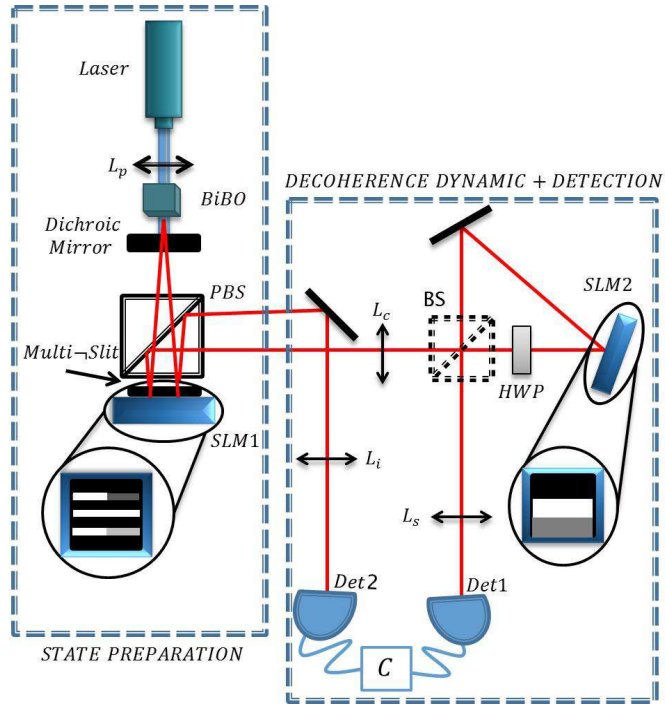}
\end{center}
\caption{Experimental setup for implementing dephasing and damping dynamics for qudits encoded in transverse path states of the twin photons. The lens $L_p$ focuses the pumping beam at the multi-slit plane, generating the state given by equation \ref{expstate} after the multiple slit \cite{TRANSVERSAL1}. The SLM1, together with a PBS and the multi-slit can generate partially entangled states. To perform the dephasing dynamics, a spherical lens $L_c$ is placed in the configuration $2f-2f$, creating a multi-slit image on the SLM2 plane. A beam splitter (BS) depicted in dotted line is not used for this implementation. On the amplitude damping dynamics, a cylindrical lens is used to project the image at an infinite distance, instead of the lens $L_c$. In order to realize the damping operations, we place the BS (dotted line) to construct a Sagnac interferometer. The lenses $L_i$ and $L_s$ are used to create an image or interference pattern according to their positions at the focal plane. We use a half wave plate (HWP) to rotate the polarization of signal photons because the SLM2 modules only horizontal polarization.}
\label{setup1}
\end{figure}

In this section we describe the setup for simulating experimentally dephasing and damping dynamics on qudits. The experimental setup is illustrated in Figure~\ref{setup1}. A $100$~mW solid state laser operating at $\lambda=355$~nm pumps a $5$~mm thick type~I BiBO crystal (BiB$_{3}$O$_{6}$) and creates degenerate non-collinear photon pairs with horizontal polarization. A dichroic mirror placed after the crystal reflects the pump beam out of the setup and transmits the photon pairs. Signal ($s$) and idler ($i$) photons ($\lambda_{s,i}=710$ nm) are transmitted through a polarizing beam splitter (PBS) before crossing a multi-slit array placed perpendicularly to the propagation axis of the pump beam at a distance of $250$~mm from the crystal. Taking the pump beam direction as the $z$ longitudinal axis, the multi-slit plane lies in the $x-y$ transverse plane. The slits are $0.1$~mm wide and have a center-to-center separation of $0.25$~mm. A $300$~mm focal length lens $L_{p}$ placed $50$~mm before the crystal is used for focusing the pump beam at the multi-slit array plane. When the beam waist at this plane is smaller than the separation between the slits, the spatial part of the two-photon state after crossing the aperture will be given by \cite{TRANSVERSAL2, QUDITWITNESS4, TRANSVERSAL1, QUDITSESPACOLIVRE, PROJETORFENDA03, TRANSVERSALEMARANHAMENTO, PROJETORFENDA02, QUDITWITNESS3}
\begin{equation}
\ket{\psi}=\frac{1}{\sqrt{d}}\sum_{\ell=-\ell_d}^{\ell_d}\ket{\ell}_s\ket{-\ell}_i,
\label{expstate}
\end{equation}
where $\ell_d=(d-1)/2$, $D$ is the number of slits, and $\ket{\ell}_{s}(\ket{\ell}_{i})$ is the so-called signal (idler) slit state or photon path state. At this point, a maximally path entangled state is prepared.

Now, we describe how we prepare a path state with any degree of entanglement. A Holoeye LC-R 2500 spatial light modulator, depicted as SLM1 in Figure~\ref{setup1}, is positioned just behind the multi-slit array, at a distance of $\sim$1~mm, to prevent diffraction. The SLM1 display is split horizontally into two sections, one for signal and the other for idler photon; independently, each section is addressed with a multi-slit aperture with different gray levels (see the lower inset of Figure~\ref{setup1}). A particular gray level is associated with a specific change in the polarization of the incoming photons. When the photons are reflected back by the SLM1, a PBS is used to filter them, selecting only their vertical polarization components. At the vertical output port of the PBS, the photon states get different modulation in each slit as they reach different gray levels at the SLM1 display. Thus, the SLM1 can be used to produce a partially entangled state, accordingly to the required application \cite{SLMTRANSVERSAL2}. In the following sections, we present the dynamics implementation.

\subsection{Dephasing implementation}
\label{Tdephasing}
We characterize a ququart photonic path state under the dephasing dynamics. Ququart states are prepared by placing a four-slit in front of SLM1, perpendicular to the photon pair path (Figure \ref{setup1}). In this particular dynamics, the SLM1 does not change the initial ququart state which is given by equation \ref{expstate}. On the idler arm, the idler photon passes through the lens $L_i$, with a focal length of $200$ mm, that projects the interference pattern at the detector 2 plane. On the signal arm, the signal photon passes through $L_c$, with a focal length of $125$ mm, which is positioned in the configuration $2f-2f$ with the SLM2 plane. This configuration allows to create the multi-slit image on the SLM2 screen. The SML2 model is a Hamamatsu LCOS-SLM X10468 which is used to perform the dephasing. The SLM2 display is addressed with four rectangular regions (see upper inset of Figure \ref{setup1}), each one matching a given slit $\ell$ from the four-slit array. Each gray level region modulates a phase $\phi_\ell$ and performs the dephasing operations $K_i$ independently. Under this implementation, the BS is not present. The lens $L_s$, with a focal length of $200$ mm projects the interference pattern on the detector 1 plane. Any operation required to implement the dephasing dynamics is performed by the SLM2.

\begin{figure}[!ht]
\hspace{0.1cm}
\begin{center}
\includegraphics[width=13cm]{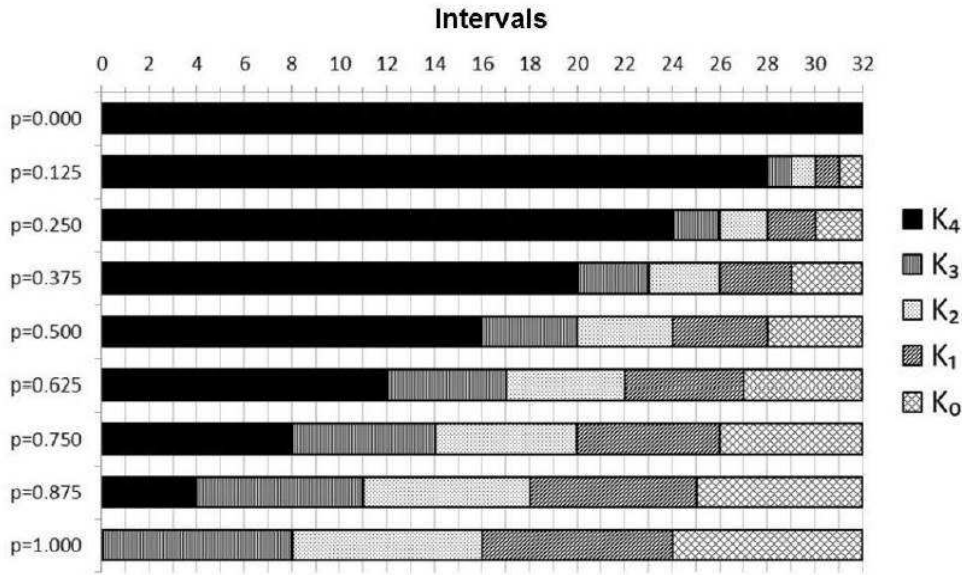}
\end{center}
\caption{The graphic shows how the films with the SLM patterns are formed to produce the dephasing dynamic. The different shades of gray represent different Kraus operators. The Kraus operators are implemented by the SLM2, which is divided in four rectangular regions corresponding to the four paths available to the signal photon in a ququart state.}
\label{intervalosdetempo}
\end{figure}

Let us consider the state $\rho_S$ that describes an ensemble with $N$ components. When the dephasing occurs, the constituent $N p_i$ evolves according to the operator $K_{i}$ (see equation \ref{pspi}), with $i= 0, 1, 2, 3, 4$. In order to implement this dynamics in our experimental setup, we explore a way to divide the ensemble described by $\rho_S$. This novel technique makes partitions over the acquisition time where different operators $K_{i}$ acts on each time division. Instead of using a single static image on the SLM2, we use a film, which are kinetic images related to the operations $K_i$. During a certain time interval which corresponds to a single kinetic image, the four rectangular regions at SLM2 will have different gray levels but constant at this time division. The gray levels are chosen such that the path phases added by the SLM2 implement the Kraus operators $K_{i}$ (eq. 8 and 9). A sequence of 32 of these gray levels patterns changing at each time division constitutes what we call a film. The time duration of a film is equal to the acquisition time. Because the SPDC process generates randomly photon pairs, on average the same number of down-converted photons are generated at equal intervals. So, our ensemble is formed by twin-photons generated over equal time intervals. 

Therefore the acquisition time is partitioned into 32 equal time intervals and the parameter $p$ is implemented over a whole acquisition time. Also, the parameter $p$ varies according to each film. In Figure~\ref{intervalosdetempo} illustrates the implementation of the dephasing dynamics for distinct values of parameter $p$. For parameter $p=0.000$, the film is composed by 32 equal consecutive images that perform 32 times the operator $K_4$ (identity). For parameter $p=0.125$, the film is composed by 28 equal consecutive images (operator $K_4$), followed by four consecutive images related to operators $K_i$, with $i= 0, 1, 2, 3$, respectively. For $p=1.000$, the film is composed by 4 sets of 8 equal consecutive images which perform the operators $K_0$, $K_{1}$, $K_{2}$ and $K_3$ exposed at equal time intervals, while the image related to operator $K_{4}$ is not included in the slides sequence. Repeating the formerly procedure, a dephasing dynamics was performed varying the parameter $p$ in steps of $0.125$ within the interval $\left[0,1\right]$.

Dephasing dynamics can be characterized through the two-photon conditional interference patterns. Such patterns allow us to obtain the experimental values for the parameters $p$. The conditional pattern on the detection plane \cite{POVMFOURIER} is described by the equation below 
\begin{eqnarray}
 P(x_{i},x_{s})=A sinc^{2}\left(\frac{kax_{i}}{f}\right)sinc^{2}\left(\frac{kax_{s}}{f\beta}\right) \nonumber\\
\times\left[1+\sum_{\ell>m=\ell_{d}}^{\ell_{d}}(1-p)|\bra{\ell,-\ell}\rho(0)\ket{m,-m}|sinc\left((\ell-m)\frac{kdb}{f}\right)sinc\left((\ell-m)\frac{kdb}{f\beta}\right)\right. \nonumber\\
\times\left.\cos\left((\ell-m)\frac{kdx_{i}}{f}-(\ell-m)\frac{kdx_{s}}{f\beta}+\arg(\bra{\ell,-\ell}\rho(0)\ket{m,-m})\right)\right],
\label{fitqubitsdefasagem}
\end{eqnarray}
where $k$ is twin-photon wave number, $2a=0.1~mm$ is the slit width, $d=0.250~mm$ is the distance between the slits, $f=200~mm$ is the  focal length of a convergent lens placed at a distance $f$ from the detectors plane, $2b=0.1~mm$ is the detectors width, $A$ is a normalization constant and $\rho(0)=\ket{\psi}\bra{\psi}$ is the initial state given by equation \ref{expstate}. The scale factor $\beta$ has a dimensionless value of $0.62$ and appears due to the combination of the lens $L_c$ and $L_s$.

\begin{table}[ht!]
\caption{\label{deftable} Experimental and predicted values of $p$ for dephasing dynamics on ququarts. Experimental parameter $p_{x=0}$ ($p_{x=x_{\pi}}$) is obtained by fitting the interference pattern curves measured by detecting the photons in coincidence, keeping fixed the signal detector (det. 1) at $x=0$ ($x=x_{\pi}$) and scanning the idler detector (det. 2). Predicted values $p_{predicted}$ are the parameters we attempt to implement.}
\begin{center}
\begin{tabular}{|c||c|c|}
\hline
$p_{x=0}$ & $p_{x=x_{\pi}}$ & $p_{predicted}$\\
\hline\hline
$0.000\pm0.063$ & $0.000\pm0.044$ & $0.000$ \\
$0.101\pm0.066$ & $0.113\pm0.055$ & $0.125$\\
$0.234\pm0.071$ & $0.248\pm0.054$ & $0.250$\\
$0.359\pm0.071$ & $0.369\pm0.052$ & $0.375$\\
$0.489\pm0.066$ & $0.490\pm0.053$ & $0.500$\\
$0.603\pm0.073$ & $0.617\pm0.058$ & $0.625$\\
$0.737\pm0.062$ & $0.753\pm0.051$ & $0.750$\\
$0.871\pm0.050$ & $0.869\pm0.061$ & $0.875$\\
$0.982\pm0.069$ & $0.94\pm0.10$ & $1.000$\\
\hline
\end{tabular}
\end{center}
\end{table}

\begin{figure}[ht!]
\hspace{0.1cm}
\begin{center}
\includegraphics[width=10cm]{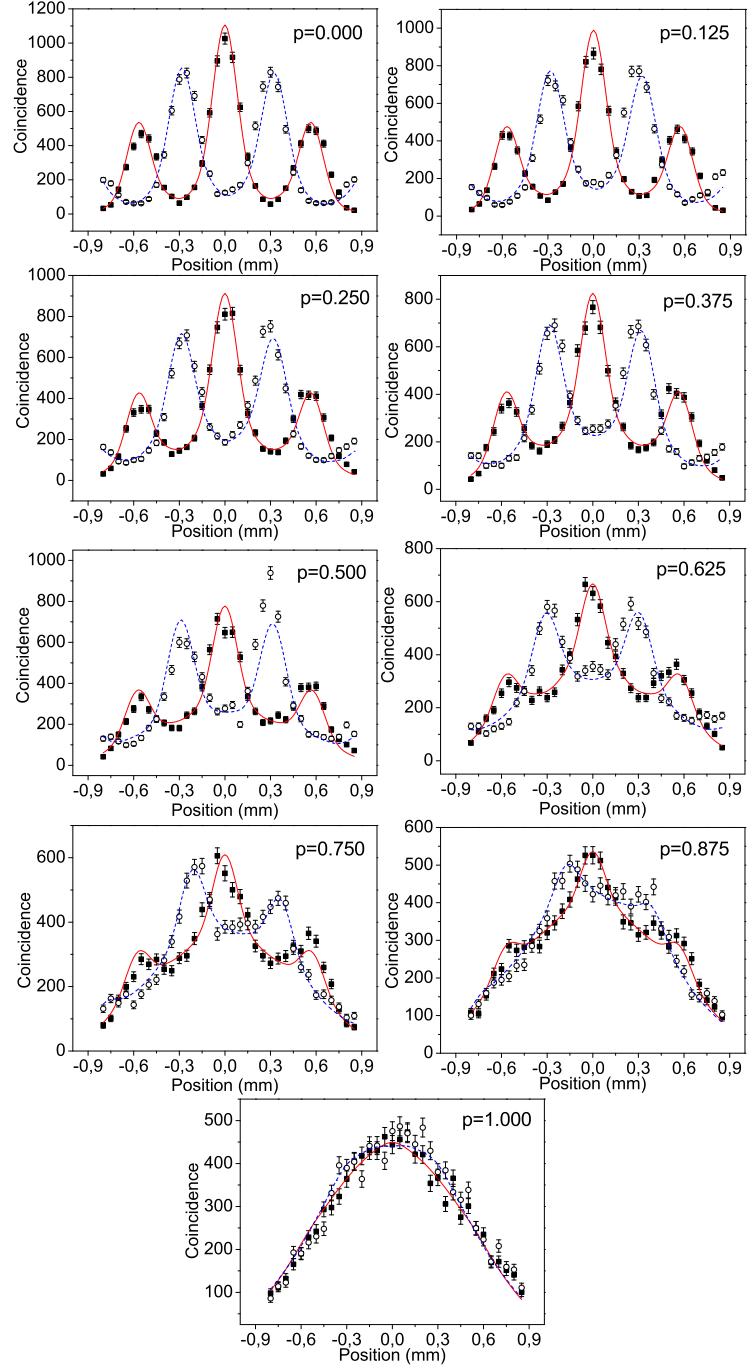}
\end{center}
\caption{Ququart interference pattern when the idler detector is scanned and the signal photon is fixed in $x_s=0$ (closed squares) and $x_s=x_\pi$ (open circles) for all $p$ values measured. To find the $p$ value for each pattern, we fitted the graphs using the equation \ref{fitqubitsdefasagem} with $D=4$.}
\label{Rdephasing}
\end{figure}

To obtain a single value of parameter $p$, we measure two interference patterns by varying the idler detector's position along the x-direction while the signal detector is positioned at $x_s=0$ and $x_s=x_\pi$ ($kdx_\pi/f\beta=\pi$). Figure \ref{Rdephasing} shows the interference patterns and their fits, which are performed by fixing all parameters, except $p$. As expected, the interference pattern visibilities decrease when the parameter $p$ increases. In Table \ref{deftable}, we report the experimental (obtained from conditional interference patterns) and the predicted (calculated from time duration of slides at the film) values of $p$.

\subsection{Amplitude damping implementation}
\label{Tamplitude}
\begin{figure}[!ht]
\hspace{0.1cm}
\begin{center}
\includegraphics[width=8cm]{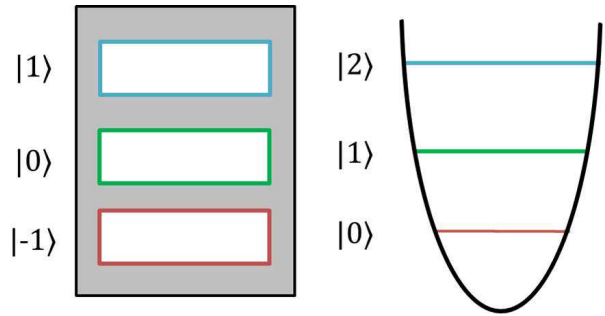}
\end{center}
\caption{Schematic representation of the correspondence between the photon path states defined by slits and the harmonic oscillator.}
\label{3niveis}
\end{figure}

The damping dynamics is performed in spatial qutrit states using a three-slit for defining the photon paths. At the SLM1, each slit region is modulated differently to get a specific amplitude and, as we mentioned above at the beginning of section \ref{setup}, we prepare a different two-qutrit state with the state coefficients modified. In this way, we prepare a partial entangled qutrit state of the form
\begin{equation}
\ket{\psi(0)}=a\ket{-1_s,1_i}+b\ket{0_s,0_i}+c\ket{1_s,-1_i},
\end{equation}
with $a^2+b^2+c^2=1$. This qutrit system behaves like a truncated harmonic oscillator, where the relation between the states $\{\ket{0}, \ket{1}, \ket{2}\}$ and the slits are shown in Figure \ref{3niveis}.
The no-jump trajectory of this amplitude damping dynamic  is implemented only in the path states of the signal photon and the two-qutrit state evolution is described by
\begin{eqnarray}
\ket{\psi(t)}&=&e^{-\frac{i}{\hbar}(H_{eff}\otimes I)t} \ket{\psi(0)}\\
&=&\frac{1}{N(t)}\left(a\ket{0_s2_i}+be^{-\gamma t}\ket{1_s1_i}+ce^{-2\gamma t}\ket{2_s0_i}\right),\nonumber
\label{dampingstate}
\end{eqnarray}
where $N(t)$ is a normalization factor and $\gamma$ is the analogous constant decay rate present in the treatment of cavity loss.

To implement the no-jump operations, we placed a cylindrical $L_c$ lens for projecting the image of the three-slit array at infinity. This image is propagated along the signal arm to detector 1, passing into a Sagnac interferometer whose input and output ports are defined by a $50/50$ beam-splitter (dotted BS showed in Figure \ref{setup1}). Inside the Sagnac interferometer there is a HWP which changes vertical polarization into horizontal and vice-versa. This interferometer introduces phase differences ($\phi_\ell$) between transmitted and reflected paths inside the Sagnac interferometer since the SLM2 modulates phases only for horizontally polarized photons. So, this modified Sagnac interferometer performs modulation in each slit state described by the unitary operation
\begin{equation}
U_{sag}\ket{\ell_s}=\sin\left(\frac{\phi_\ell}{2}\right)\ket{\ell_s}.
\label{sagnacoperation}
\end{equation}

\begin{figure}[!ht]
\hspace{0.1cm}
\begin{center}
\includegraphics[width=10cm]{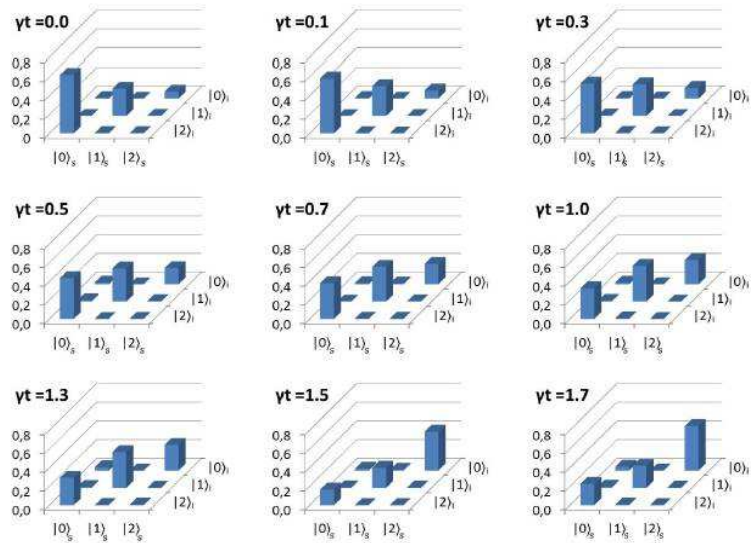}
\end{center}
\caption{System population measurement over time. As expected, the slit populations related to higher energy level decrease and the lower energy levels increase.}
\label{amplitudepopulacao}
\end{figure}

Comparing the above Sagnac operation expression with the state evolution (equation \ref{dampingstate}), it is possible to find a correspondence between to analogous physical systems: the photon decay rate inside a cavity and the phase modulation given by the interferometer (figure \ref{setup1}). Therefore, the amplitude damping (no-jump) dynamics can be performed by introducing the following phase differences between the reflected and transmitted photon path states at the interferometer
\begin{eqnarray}
\phi_0(t)=0,\nonumber\\ \phi_1(t)=\arcsin\left(e^{-\gamma t}\right), \\ \phi_2(t)=\arcsin\left(e^{-2\gamma t}\right)\nonumber.
\end{eqnarray}
State evolution is obtained by detecting coincidence counts of the three-slit aperture at the image plane. On the idler arm, a lens $L_i$ is used to project the three-slit image at the plane of detector 2. Both detectors 1 and 2, considered as point-like detectors Figure \ref{setup1}, are fixed at the positions corresponding to the slit image $l_s$ and $m_i$ (with $l,m=-1,0,1$), respectively. Coincidence counts are recorded for the nine measurement combinations, and by normalizing them we obtain the absolute values of amplitudes for all nine states of the slit-state basis $\{\ket{l}_s\ket{m}_i\}$. The experimental results from the measurements described above are shown in Table \ref{tabelaamplitude} and Figure \ref{amplitudepopulacao}. Such measurements alone can not fully characterize the state, however they do show the population change which is the main effect generated by the damping dynamics. 

\begin{figure}[!ht]
\hspace{0.1cm}
\begin{center}
\includegraphics[width=10cm]{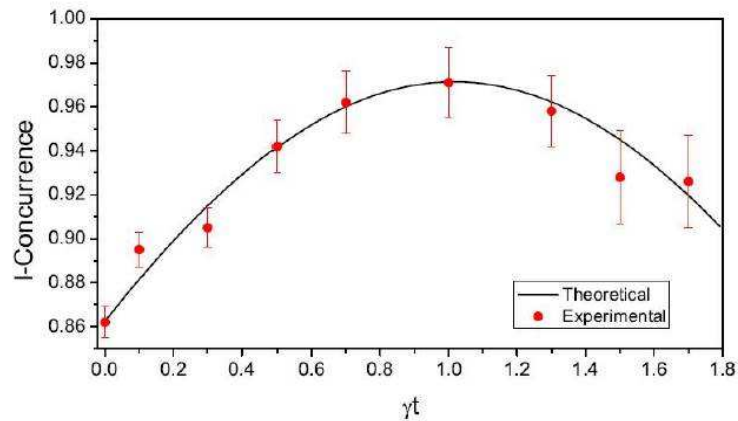}
\end{center}
\caption{Entanglement dynamics for no-jump trajectory of amplitude damping acting on the qutrit system. I-concurrence is calculated from experimental data (red dots) showed in the table \ref{tabelaamplitude} and the theoretical prediction (black line).}
\label{amplitudeimagem}
\end{figure}

Furthermore, another interesting feature of this dynamics is the increase of entanglement in the system according to the initial state. In this implementation, we generate a initial state with the constraint ($a<b<c$). In Figure \ref{amplitudeimagem} it is shown the entanglement dynamics of the system. To quantify the entanglement we used the normalized I-concurrence~\cite{rungta2001universal} defined as
\begin{equation}
C(\psi(t))=\frac{1}{\Omega}\sqrt{2\left[1-Tr(\rho_i^2)\right]}=\frac{1}{\Omega}\sqrt{2\left[1-Tr(\rho_s^2)\right]},
\end{equation}
where $\rho_i$ ($\rho_s$) is the reduced state of idler (signal) and $\Omega=\sqrt{2(d-1)/d}$. Note that the experimental error of the I-concurrence calculated from the experimental measurements increases over time. This happens due to the decrease in coincidence counts, as shown in Table \ref{tabelaamplitude}. From the above results we can infer from the decrease of the total coincidence counts that when the system evolves fewer ensemble components have no jump trajectories, as expected.

\begin{table}
 \renewcommand{\arraystretch}{1.5}
 \caption{\label{tabelaamplitude} Coincidence counts between photons transmitted by different slit states for different values of $\gamma t$. The I-concurrence $C_e(\ket{\psi(\gamma t)})$ is calculated from the measured states and compared with the predicted I-concurrence $C_p(\ket{\psi(\gamma t)})$. }
{\scriptsize \begin{center}
\begin{tabular}{|c||c|c|c|c|c|c|c|c|c|}
\hline
 & $\gamma t=0.0$ & $\gamma t=0.1$ & $\gamma t=0.3$ & $\gamma t=0.5$ & $\gamma t=0.7$ & $\gamma t=1.0$ & $\gamma t=1.3$ & $\gamma t=1.5$ & $\gamma t=1.7$\\
\hline\hline
$\ket{0,0}$ & $24\pm5$ & $21\pm5$ & $19\pm4$ & $28\pm5$ & $16\pm4$ & $17\pm4$ & $33\pm6$ & $12\pm3$ & $22\pm5$ \\
$\ket{0,1}$ & $0\pm0$ & $2\pm1$ & $6\pm3$ & $1\pm1$ & $10\pm3$ & $1\pm1$ & $0\pm0$ & $4\pm2$ & $0\pm0$\\
$\ket{0,2}$ & $249\pm16$ & $220\pm15$ & $262\pm16$ & $252\pm16$ & $273\pm17$ & $232\pm16$ & $248\pm16$ & $245\pm16$ & $252\pm16$\\
\hline
$\ket{1,0}$ & $4\pm2$ & $0\pm0$ & $12\pm3$ & $20\pm4$ & $1\pm1$ & $6\pm2$ & $9\pm3$ & $3\pm2$ & $7\pm3$ \\
$\ket{1,1}$ & $953\pm31$ & $814\pm29$ & $775\pm28$ & $517\pm23$ & $465\pm22$ & $339\pm18$ & $345\pm19$ & $227\pm15$ & $125\pm11$\\
$\ket{1,2}$ & $10\pm3$ & $4\pm2$ & $1\pm1$ & $7\pm3$ & $4\pm2$ & $0\pm0$ & $0\pm0$ & $2\pm1$ & $1\pm1$\\
\hline
$\ket{2,0}$ & $2042\pm45$ & $1490\pm39$ & $1222\pm35$ & $635\pm23$ & $476\pm22$ & $290\pm18$ & $265\pm16$ & $98\pm10$ & $120\pm11$ \\
$\ket{2,1}$ & $10\pm3$ & $4\pm2$ & $11\pm3$ & $0\pm0$ & $3\pm2$ & $5\pm2$ & $1\pm1$ & $0\pm0$ & $3\pm1$\\
$\ket{2,2}$ & $11\pm3$ & $0\pm0$ & $5\pm2$ & $2\pm1$ & $2\pm1$ & $0\pm0$ & $4\pm2$ & $0\pm0$ & $0\pm0$\\
\hline
\hline
$C_{e}(\ket{\psi(\gamma t)})$ & $0.862\pm0.007$ & $0.895\pm0.008$ & $0.905\pm0.009$ & $0.942\pm0.012$ & $0.962\pm0.014$ & $0.971\pm0.016$ & $0.958\pm0.016$ & $0.928\pm0.021$ & $0.926\pm0.021$ \\
\hline
$C_{p}(\ket{\psi(\gamma t)})$ & $0.864$ & $0.881$ & $0.914$ & $0.942$ & $0.960$ & $0.972$ & $0.962$ & $0.945$ & $0.919$\\
\hline

\end{tabular}
\end{center}}
\end{table}

\section*{Summary}
\label{Summary}

In this work we experimentally demonstrated simulations of dissipative dynamics on quantum systems in a simple implementation. Our quantum systems are spatial qudits encoded in the transverse paths of photons pairs generated by SPDC. The dissipative operators for simulating dephasing and amplitude damping dynamics were realized by means of a spatial light modulator. Dephasing dynamics was performed completely and amplitude damping dynamics was implemented partially, in which we performed only the no-jump trajectory. In the dephasing dynamics we measured the interference patterns to calculate the parameter $p$, related to coherence loss. Besides, in the amplitude damping dynamics we measured the population which allows us to identify population changes from higher to lower levels. To identify the implementation success we used a parameter that represents the principal characteristic of the evolution: coherence loss for dephasing and population changes for amplitude damping. Spatial photonic states are interesting systems to make quantum computation using qudits. The implemented dissipative dynamics are general and do not depend on the initial state and can also be extended to different system dimensions. Moreover, the experimental technique using films instead of images can be manipulated to implement other types of operations, increasing the SLM uses. 

\section*{Acknowledgments}
This work is part of Brazilian National Institute for Science and Technology for Quantum Information and was supported by the Brazilian agencies CNPq, CAPES, and FAPEMIG. We acknowledge the EnLight group for very useful discussions.


\end{document}